\documentstyle[12pt]{article}
\catcode`@=11   
\let\l=\lambda

\def\IR{\relax\leavevmode{\rm I\kern-.18em R}}
\def\ZZ{\relax\leavevmode
       \ifmmode\mathchoice
       {\hbox{\sf Z\kern-.4em Z}}
       {\hbox{\sf Z\kern-.4em Z}}
       {\lower.9pt\hbox{\scriptsize\sf Z\kern-.36em Z}}
       {\lower1.2pt\hbox{\tiny\sf Z\kern-.36em Z}}
       \else{\sf Z\kern-.4em Z}\fi}
\def\RR{\relax\leavevmode
       \ifmmode\mathchoice
       {\hbox{\sf R\kern-.4em R}}
       {\hbox{\sf R\kern-.4em R}}
       {\lower.9pt\hbox{\scriptsize\sf R\kern-.36em R}}
       {\lower1.2pt\hbox{\tiny\sf R\kern-.36em R}}
       \else{\sf R\kern-.4em R}\fi}

\def\resetby#1#2{\@addtoreset{#2}{#1}}
\def\seceq{\@addtoreset{equation}{section}
              \def\theequation{\thesection.\arabic{equation}}}

\def\Label#1{\label{#1}%
                \smash{\hbox to0pt{\raise1ex\hbox{\tiny[#1]}\hss}}}
\def\noLabels{\let\Label=\label}

\thicklines     

\reversemarginpar
\catcode`@=12

\begin{document}

\begin{titlepage}

\begin{flushright}

VPI-IPPAP-03-08\\

hep-th/0305193\\

\end{flushright}

\begin{center}

\vskip 1cm

{\large \bf
          {Background Independent Quantum Mechanics and
Gravity}}\\[5mm]
{\bf Djordje~Minic\footnote{e-mail: dminic@vt.edu}
and Chia-Hsiung Tze\footnote{e-mail: kahong@vt.edu}} \\[1mm]
            Institute for Particle Physics and Astrophysics\\
            Department of Physics\\
            Virginia Tech\\
            Blacksburg, VA 24061\\[5mm]

\vskip 1cm

{\bf ABSTRACT}\\[3mm]

\parbox{4.9in}{We argue that the demand of
background independence in a quantum theory of gravity calls for an extension of standard geometric
quantum mechanics. 
We discuss a possible kinematical and dynamical
generalization of the latter by way of a quantum covariance of the state space.
Specifically, we apply our scheme to the problem of a background
independent formulation of Matrix Theory.}

\end{center}

\end{titlepage}

The quest for a consistent, unified quantum theory of matter and gravity remains very much an open issue, despite great progress in 
string theory
\cite{string} and other approaches to quantum gravity \cite{qgrav}. At the outset either quantum mechanics (QM) or general relativity (GR) or both should give way to a new substructure.  From the predominant viewpoint in string theory it is GR that needs replacing, while QM is complete by itself.  This stand is well motivated
when GR is taken as valid only at low energy
scales \cite{string}. In fact, the most general diffeomorphism invariant effective
action is {\it derivable} from the consistency requirements (i.e. conformal invariance) of a perturbative string theory. Alas, a non-perturbative, background independent understanding
of this remarkable fact is still lacking.
Most of other attempts at quantizing gravity
\cite{qgrav}, while stressing the relational, background independent nature
of GR, (and thus arguing against the effective field theoretic point of view and
for a non-standard approach to quantization of gravity), still keep to
the canonical structure of QM\footnote{For a recent
critical discussion on the status of canonical QM in
quantum gravity, see, for example \cite{isham}.}.

What is, may one ask, the physical rationale for changing the canonical
quantum mechanical structure? And if such a rationale is found,
is it theoretically/empirically compelling?
In the negative, isn't the canonical structure
somehow unique, and if so, what does this imply for the foundational
issues of quantum theory, of a theory of quantum gravity in particular?

This letter puts forth a radical but, in our view, a
justified approach to extending QM. Motivated by the physical requirement of background independence (BI) and the need to make room for gravity at the quantum level, we are led to a rather drastic extension of standard QM: to wit, we modify both its dynamics and kinematics, and thereby the very symplectic and Riemannian structures that underlie its geometric foundations\footnote{Our
present discussion is far more general than our recent  
geometric formulation of Nambu quantum 
mechanics \cite{tzeminic}, motivated by the covariance problem in Matrix theory.}.
{\it The upshot of our proposal is that the space of quantum states (events)
becomes {\it dynamical} and that the dynamical geometric information
is described in terms of a non-linear diffeomorphism invariant theory, in such a way that
the space of quantum events is non-linearly inter-related
with the generator of quantum dynamics - the Hamiltonian.}

We briefly review the key features of geometric standard QM \cite{kibble} (for
reviews of this approach consult \cite{weinberg}, \cite{reviews}, \cite{Bloch}, \cite{anandan}, \cite{provost}).   Pure states are points of an infinite dimensional Kahler manifold  ${\cal{P(H)}}$, the complex projective space of  the Hilbert space ${\cal{H}}$. Equivalently, ${\cal{P(H)}}$ is a real manifold with an integrable almost complex structure $J$. As such it has a Kahler metric given by ${<\psi|\phi>}$,  the Hermitian inner product of two states $<\psi|$
and $|\phi>$ in ${\cal{H}}$, the Riemannian metric $g(\psi, \phi)$ = $g(J\psi, J \phi)$ = 2 $k  Re(<\psi|\phi>)$ which is uniquely the Cayley-Fubini-Study metric.  The associated symplectic 2-form $\omega(\psi,\phi)$ = 2 $k  Im(<\psi|\phi>)$  with $k$ = $\hbar$ = 2/c  with $h$ being Planck's constant and $c$ the constant holomorphic sectional curvature (CHSC) of ${\cal{P(H)}}$.
While the role of the symplectic
structure is well known, the Riemannian structure, notably absent in classical phase space, is the key player and as the metric structure on ${\cal{P}}$
embodies the information about
purely quantum properties, such as time evolution,
 uncertainty relations, entanglement and the measurement process.
The correspondence with the operatorial formalism is as follows:  An observable
${A = <\hat{A}>}$,
defined as the expectation value of a hermitian linear operator $\hat{A}$,
is a real valued
differentiable function on ${\cal{P}}$, belonging a special class of $Kahlerian functions$. This contrasts sharply with classical Hamiltonian dynamics where {\it any} function of the canonical variables is an observable \footnote{Such a possibility in the context of
a generalized QM was analyzed by Weinberg \cite{weinberg}.}.  The derivative of such a
 $A$ vanishes
at an ``eigenstate'' with the value of $A$ at such a point giving the
``eigenvalue''.

The evolution of states i.e. the Schr\"{o}dinger  equation is given by the
symplectic flow generated
by a Hamiltonian $H$ of any given system.
Let a pure state be $ \psi = \sum_a e_a \psi_a$, where the
$\psi_a$ are the coefficients of $ \psi $ in an
orthonormal eigenbasis
$\{e_a\}$ of $H$. Let
$q^a =
\sqrt{2\hbar} Re \psi_a$ and
$p_a = \sqrt{2\hbar} Im \psi_a$ with the $(q^a + ip_a)$ being the
homogeneous coordinates
for  ${\cal{P}}$. The symplectic structure on ${\cal{P}}$ is
given by the closed, nondegenerate 2-form
$\omega^{(2)}= d p_a \wedge dq^a$, $d\omega^{(2)} = 0$. The Poisson bracket is defined as usual:
$
\{f, g\} = {\partial f \over \partial p_a }
{\partial g \over \partial q^a} -
{\partial f \over \partial q^a}
{\partial g \over \partial p_a} \equiv \omega^{ab}  {\partial f \over
\partial X^a}
{\partial g \over \partial X^b},
$
where $ \omega^{ab}$  is the inverse of  $\omega^{(2)}$ and the $X^a
= (p_a, q^a)$ form a set of canonical coordinates.  The
Schr\"{o}dinger equation, with ${h = <\hat{H}>}$, is then simply Hamilton's equations:
$
{d p_a \over dt} = \{h, p_a \}, \quad {d q^a \over dt} = \{h, q^a\}.
$
Here $h = {1 \over 2} \sum_a [ (p^a)^2 + (q_a)^2 ] \omega_a$, $\omega_a$ being the energy eigenvalues.
An observable $O$ will then evolve as
$
{ d O \over dt}= \{h, O \}.
$
The expectation values of commutators of
operators acting on
${\cal{H}}$ are precisely the Poisson brackets of the corresponding Kahlerian functions!
The inner product determines the flat metric on the Hilbert space
$
dS^2 = \sum [(dq_a)^2 + (dp_a)^2] = \delta_{ab} dX^a dX^b
$
where $X^a = (q^a, p^a)$.

The Born rule $\psi^{*} \psi =1$
= ${1 \over 2\hbar} \sum_a [ (p^a)^2 + (q_a)^2 ] =1 $ implies that
$\psi$ and
$e^{i\alpha} \psi$
are to be identified. For finite $n$, 
we then have
as the space of rays in ${\cal{H}} = C^{n+1}$,
the complex
projective space $CP(n)$, the base space of the complex
Hopf line bundle of the sphere
${S^{2n+1}}$ over $CP(n)$ = $U(n+1)/
U(n) \times U(1)$) with its $U(1)$ fiber, the group of complex phases in
QM. Thus QM can be viewed as a classical Hamiltonian system, albeit a very special one with, as its phase space, the nonlinear, rich and "huge"
projective Hilbert space $CP(n)$ with $n = \infty$ generically
\cite{Bloch}
and $U(n)$ as the unitary group of quantum canonical transformations.
The unique Riemannian metric on $CP(n)$, induced from the inner product of  ${\cal{H}}$, is the Fubini-Study (FS) metric, $ds_{12}^2 = (1 -
|\langle \psi_1|\psi_2 \rangle|^2)\equiv 4(\langle d\psi|d\psi\rangle
- \langle d\psi|\psi\rangle\langle \psi|d\psi\rangle)$.
The Heisenberg uncertainty relations arise from such a
metric  of $CP(n)$ whose local properties also yield a generalized
energy-time uncertainty relation
\cite{anandan}.  The probabilistic (statistical) interpretation of
QM is thus hidden in the metric properties of ${\cal{P(H)}}$. The unitary time evolution is
related to the metrical structure \cite{anandan} with Schr\"{o}dinger's equation in the guise of a geodesic equation
on  $CP(n)$ = $U(n+1)/U(n) \times U(1)$
$:
{d u^a \over d s} + \Gamma^{a}_{bc} u^b u^c =
\frac{1}{2\Delta E}Tr(H F^a_b) u^b
$
for the FS metric $g_{ab}^{FS}$
with the canonical curvature 2-form $F_{ab}$ valued in the holonomy gauge group $U(n)\times U(1)$. Here $\Delta E^2 = \langle H^2 \rangle - \langle H \rangle ^2$ as
in \cite{anandan}.
Also, $u^a = \frac{d z^a}{ds}$ where $z^a$ denote the complex
coordinates on $CP(n)$ and 
$\Gamma^{a}_{bc}$ is the connection obtained from the FS metric.
The affine parameter $s$ is determined by the $CP(n)$ 
metric.
As underscored by Aharonov and Anandan \cite{anandan},  time measurement in the evolution of a given system
reduces to that of distance on $CP(n)$. In particular
$
\hbar ds = 2 \Delta E dt.
$ 
Such an expression naturally invokes
a {\it{relational}} interpretation of time in QM. Even more striking is the fact that the geometric interpretation of
probability as the geodesic distance on  $CP(n)$
is {\it{directly}} related to the definition of
the evolution parameter $t$!
In the above geodesic Schrodinger equation,
$H$ appears as the ``charge'' of an
effective particle moving with a ``velocity'' $u^a$ in the background
of the ``Yang-Mills'' field $F_{ab}$.
Finally, given a curve $\Gamma$ in the projective Hilbert space
${\cal{P}}$, the geometric (Berry) phase \cite{gphase}
is given by \cite{anandan}
$
\int _{\Sigma} d p_a \wedge dq^a,
$
where $\Sigma$ has as its boundary $\Gamma$. As a
symplectic area enclosed by $\Sigma$, this phase depends
solely on the geometry of the inner product and is both
independent of the Hamiltonian and the equation of motion iff the latter is first order in time.

Next we draw attention to a simple calculation of 
$ds_{12}^2 $
for the Gaussian coherent state
$
\psi_l(x) \sim \exp(- \frac{{({\vec{x}}-{\vec{l}})}^2}{\delta l^2})
$
which, using the convolution property of Gaussian integrals, yields the natural metric in the configuration space, namely
$
ds^2 = \frac{d {\vec{l}}^2}{\delta l^2}.
$
So, wherever the configuration space coincides with 
{\it space}, the natural metric on $CP(n)$
in the $\hbar \to 0$ limit gives a spatial metric \cite{anandan}.
{\it It is this important insight which is the springboard for our proposed background
independent generalization of standard QM.} 
For a generalized coherent state, the
FS metric reduces to the metric on the corresponding group manifold
\cite{provost}.

Given the Riemannian structure of QM and the observed connection between the FS and the spatial metrics, it behooves us to inquire if a more general Riemannian
structure of space can be induced from a more flexible
state space than $CP(n)$. Specifically, let $l$ in the
above metric computation be mapped to
$\l \to k(l)$.
The corresponding expression for the spatial metric results from
the overlap of two Gaussians
$
\psi_{k(l)}(x) \sim \exp(- \frac{(x-k(l))^2}{\delta k(l)^2})
$
which in turn follows
from
$
\int dx g_{\psi_l, \psi_{l+dl}} \psi_l^* \psi_{l+dl}
$
where the ``quantum metric'' reads
$
g_{\psi_l, \psi_{l+dl}} \equiv \frac{ \psi_k^* \psi_{k+dk}}{ \psi_l^* \psi_{l+dl}}.
$
Clearly the transformation that takes $\psi_l \to \psi_{k(l)}$ is $not$ in general unitary.
If we insist on the desired relation between the quantum metric 
and an arbitrary metric on the classical configuration space, then the kinematics of QM must
be altered.  Moreover if the induced classical configuration space is to be the
actual space of spacetime, only a special
quantum system will do.  We are thus induced to make the state manifold suitably flexible by doing general relativity on it. The resultant metric on the Hilbert space is
generally curved with its
distance function modified, an extended
 Born rule and hence a new meaning to probability.
By insisting on diffeomorphism invariance in
the state space and on preserving the desirable complex projective properties of Cartan's rank one symmetric spaces such as $CP(n)$, we arrive at the ensuing coset state space
$Diff(\infty, C)/Diff(\infty -1,C) \times Diff(1,C)$ as the minimal phase space candidate for a background independent QM.
In summary, the axioms of standard geometric QM are enlarged as follows:

1) The state space  $CP(\infty$) is extended to
$Diff(\infty, C)/Diff(\infty -1) \times Diff(1,C)$ deriving from the generalized
inner product
\begin{equation}
dS^2 = \sum h_{ab}[(dq_a)^2 + (dp_a)^2] \equiv h_{ab} dX^a dX^b,
\end{equation}
where $h_{ab}$ is hermitian.
The``Born rule'' now reads
\begin{equation}
{1 \over 2} \sum_{a,b} h_{ab} [ (p^a p^b) + (q^a q^b) ] =1 .
\end{equation}
These equations provide the metric relation on and the geometrical shape(s) of the new
state space, and implicitly defines $\hbar$.
The probabilistic interpretation lies in the
definition of geodesic length on the new space of quantum states (events).
The relation $\hbar ds = 2 \Delta E dt$ gives 
meaning to the ``evolution parameter'' $t$!
Notably different metrics imply different
``evolution parameters'' with $t$ relational and akin to
the ``multifinger time'' of GR \cite{mtw}.
Given the $X$ space, we can introduce a natural $Diff(1,C)$ map, $X \to f(X)$.
The $Diff(1,C)$ identification of the points on the submanifold
determined by the " Born rule"
defines the generalized projective Hilbert manifold. 

2) The observables are functions of the natural distance
on the quantum phase space
$h_{ab} X^{a} X^{b}$, $O = O(h_{ab} X^{a} X^{b})$.
They reduce to the usual ones when 
the Riemannian structure is canonical. More explicitly
\begin{equation}
O = \sum_{a,b} o_{a} h_{ab} X^a X^b
\end{equation}
where the ``eigenvalue'' $o_a$ is given as (see \cite{weinberg})
\begin{equation}
{ d O \over d X^a}= o_a \omega_{ab} X^b.
\end{equation}
Here the symplectic form $\omega^{ab}$ as well as $O$ depend 
on the invariant combination $h_{ab} X^{a} X^{b}$.

3) The temporal
evolution equation reads
\begin{equation}
{d u^a \over d\tau} + \Gamma^{a}_{bc} u^b u^c =
\frac{1}{2 \Delta E}Tr(H F^a_b) u^b
\end{equation}
where now $\tau$ is given through the metric
$\hbar d\tau = 2 \Delta E dt$,
as in the original work of Aharonov and Anandan \cite{anandan}.
$\Gamma^{a}_{bc}$ is the affine connection associated with this general metric 
$g_{ab}$ and $F_{ab}$ is a general curvature 2-form in 
$Diff(\infty -1, C) \times Diff(1,C)$.

Next we reformulate geometric QM in the above background independent
(BI) setting. Due to
the $Diff(\infty, C)$ symmetry, ``coordinates'' $z^a$ (i.e. quantum states) make no sense physically,
only quantum events do, which is the quantum counterpart of the
corresponding statement on the meaning of space-time events in 
GR.
Probability is generalized and given by the notion of diffeomorphism invariant distance in the space of quantum configurations.
The dynamical equation is a geodesic equation on
this space. Time, the evolution parameter in the generalized Schr\"{o}dinger equation,
is not global and is given in
terms of the invariant distance. 
Our basic starting point of a background independent QM (BIQM) is to notice that the evolution equation (the
generalized Schr\"{o}dinger equation) as a
geodesic equation, can be derived from an
Einstein-like equation with the energy-momentum
tensor determined by the holonomic non-abelian field strength $F_{ab}$ 
of the $Diff(\infty -1,C) \times Diff(1,C)$ type
and the interpretation of the Hamiltonian as a ``charge''.
Such an extrapolation is logical since $CP(n)$ is an Einstein space; its metric obeying
Einstein's equation with a positive cosmological constant given
by $\hbar$:
$R_{ab} - \frac{1}{2} g_{ab} R  - \lambda g_{ab}=0$.
The Ricci curvature of $CP(n)$ is 
$R_{ab} \equiv \frac{n+1}{\hbar} g_{ab} = \frac{1}{2} c (n+1) g_{ab}$,
where $c$ is the CHSC of $CP(n)$ given by
$c=\frac{2}{\hbar}$.

The geodesic equation (5) follows from the conservation of the
energy-momentum tensor 
$
\nabla_a T^{ab} =0
$
with 
$
T_{ab} = Tr(F^{ac}g_{cd}F^{cb} -\frac{1}{4} g_{ab} F_{cd}F^{cd}
+ \frac{1}{2\Delta E}H u_a u_b)
$
by way of the usual GR argument (e.g. \cite{mtw}, chapter
20).
With quantum gravity in mind, we
 set $\Delta E$ to the Planck energy $E_p$, the proper deformation parameter.
When $E_p \to \infty$ we recover the usual flat
metric on the Hilbert space or the FS metric
on the projective Hilbert space.
Since both the metrical and symplectic data are also contained in
$H$, we have here the advertised non-linear ``bootstrap'' between the space of
quantum events and the dynamics.
The diffeomorphism invariance of the new phase space
suggests the following dynamical scheme for
the BIQM:
\begin{equation}
\label{BIQM1}
R_{ab} - \frac{1}{2} g_{ab} R  - \lambda g_{ab}= T_{ab}
\end{equation}
with $T_{ab}$ given as above (as determined by 
$F_{ab}$ and the Hamiltonian (``charge'') $H$).
Furthermore
\begin{equation}
\label{BIQM2}
\nabla_a F^{ab} = \frac{1}{2\Delta E} H u^b.
\end{equation}
 The last two equations imply via the Bianchi identity a conserved
energy-momentum tensor, $\nabla_a T^{ab} =0$ .  The latter, taken
together with the conserved ``current'' $j^b \equiv \frac{1}{2\Delta E} H u^b$,
i.e. $\nabla_a j^a =0$,
implies the generalized geodesic Schr\"{o}dinger equation.
So (\ref{BIQM1}) and (\ref{BIQM2}), being a closed system of equations for
the metric and symplectic form on the space of events,
define our BIQM.
We emphasize once again that in the limit
$E_p \to \infty$  we recover the usual structure of linear
QM. Moreover this limit does not affect
the geodesic equation
${d u^a \over d\tau} + \Gamma^{a}_{bc} u^b u^c =
\frac{1}{2 \Delta E}Tr(H F^a_b) u^b$
due to the relation
$\hbar d\tau = 2 \Delta E dt$. As such our formulation offers a tantalizing
non-linear linkage between the
metric and symplectic data embodied in $H$
and the quantum metric and symplectic data. 
The space of quantum events is $dynamical$ paralleling the dynamical role of spacetime in GR, as opposed
to the rigid, absolute state space of standard QM. This is then, in our view, the price of quantum background independence. 
To draw more concrete consequences of this kinematics made dynamical, we next specify a quantum system with its $H$. The configuration space of the quantum metric defines
a ``superspace'' (as in canonical GR \cite{super}) and the dynamics on it presumably select
a particular background.

We now demand that the configuration
space metric be the actual physical {\it spatial} metric. The suitable quantum system must then have
a very special configuration space and
should describe a quantum theory of gravity. Specifically, we seek a canonical QM of a non-perturbative form of
quantum gravity in a fixed background, with a well
defined perturbative limit and a configuration space
being the actual space. The only example we know fulfilling these criteria is 
Matrix theory \cite{matrix}. (The latter is also ``holographic'' \cite{holog}, in the sense of mean-field theory.\footnote{The relationship between holography,
unitarity and diffeomorphism invariance was explored in
\cite{rob}.}) As with other roads to quantum gravity, Matrix theory which leaves QM intact, suffers from the problem of background dependence \cite{seiberg}, \cite{mback}.\footnote{We should mention here that different arguments for revising quantum mechanics
in the framework of quantum gravity have been advanced
for example in \cite{thooft}, \cite{penrose} and \cite{holobdbm}.}

In implementing our  scheme,
we assume that the metric on the transverse space
is encoded in the metric on the quantum state space.
Then we take the Matrix theory Hamiltonian in an arbitrary background
and insert it into the defining equations of the above BIQM.
The evolution of our system then reads
$
{d u^a \over d\tau} + \Gamma^{a}_{bc} u^b u^c = \frac{1}{2 E_p}
H_M F^a_b u^b
$
where $H_M$ is the Matrix Hamiltonian ($i,j$ denote the
transverse space indices ($i=1,...,9$), $R$ is the extent of the longitudinal 11th direction)
\begin{equation}
H_M = R Tr(\frac{1}{2} P^i P^j G_{ij}(Y) + 
\frac{1}{4}[Y^i, Y^l][Y^k, Y^j] G_{ij}(Y) G_{lk}(Y)) + fermions.
\end{equation}
Here $P^i$
is the conjugate momentum to $Y^i$ ($N\times N$ hermitian matrices) given a symplectic form $\omega$.
(We adopt the symmetric ordering of matrices, see \cite{mback}.)
{\it Given this expression for $H_M$ the general equations (\ref{BIQM1}) 
and (\ref{BIQM2})
then define a background independent Matrix theory (BIMT).}
Note that in (\ref{BIQM1}) 
and (\ref{BIQM2})
$a,b$ denote the indices on the quantum space of states, whose span is
determined by the dimension of the Hilbert space of Matrix theory, given
in terms of $N$.

So the time of BIMT is manifestly not global, but is
defined by the invariant distance on the space of quantum events.
The light-front (light-cone) $SO(9)$ symmetry 
is only ``local'' (in the sense
of the generalized quantum phase space).
SUSY is generally broken since generically, we
have no background which admits globally defined supercharges. Only
``locally'' (again in the sense of the generalized quantum space)
may we talk about the correspondence
between the moduli space of the Matrix theory SUSYQM 
and the transverse space \cite{matrix}.

As to the longitudinal coordinate,
(and longitudinal momentum, given in terms of $N/R$ \cite{matrix}), they can be
made dynamical in our proposal.
The rank $N$ of the matrices implicitly defines the size of the
Hilbert space, which is seemingly fixed (the dimension of the index space is fixed.)
On the other hand, one of the fundamental features of Matrix theory is that of
being automatically second quantized; it encodes the Fock space $\{n_k \}$ in terms of
block diagonal $n_k \times n_k$ matrices \cite{matrix}.
Taking cue from this defining feature, we promote the 
points on the quantum phase space into hermitian matrices.
This is the final ingredient in our proposal.
In practice, the $u^a$s appear as hermitian matrices in 
the defining equations
(\ref{BIQM1}) 
and (\ref{BIQM2}).
So the rank of matrix-valued non-commuting transverse coordinates $Y^i$ ($N$) is made dynamical
by turning the ``coordinates'' $z^a$ of
our background independent quantum phase space
into non-commutative objects.
The asymptotic causal structure (and thus a {\it covariant background
independent structure}) only emerges in the
Matrix theory limit \cite{matrix}, $N\to \infty$, $R \to \infty$ while keeping $N/R$ fixed.  The above defining dynamical equations (\ref{BIQM1}) 
and (\ref{BIQM2}) can also be cast
in the context of Connes' non-commutative geometry \cite{connes}.
We will discuss this topic in a separate longer publication which will elaborate the
contents of this letter \cite{futurework}.

In closing, the gist of our proposal lies in the non-linear interconnection between the
metric ($G_{ij}$) and symplectic data ($\Omega_{ij}$) contained in the Hamiltonian $H$
and the quantum metric ($g_{ab}$) and symplectic data ($\omega_{ab}$, or
equivalently, $F_{ab}$).
This non-linear connection may well explain how 
(a) different degrees of freedom are associated to different backgrounds and
(b) how the observed 4-dimensional spacetime background dynamically emerges in 
Matrix theory, the pre-geometry being the dynamical stochastic geometry of the space of events.
Furthermore we can't but ponder the fascinating possibility that the very form of 
the Matrix theory Hamiltonian $H_M$ is already encoded in the non-trivial
topological structure of the space of quantum events.  This may be so if
the latter manifold is non-simply connected and is non-commutative \footnote{This in complete analogy with the concept of ``charge without charge''
of the Einstein-Maxwell system of equations in vacuum, as discussed by
Misner and Wheeler \cite{mw}.}.

{\bf Acknowledgments:}
We are indebted to V.~Balasubramanian, I.~Bars, J.~de~Boer, 
L.~N.~Chang, L.~Freidel, J.~Gates,
E.~Gimon, M.~G\"{u}naydin, V.~Jejjala, D.~Kabat, N.~Kaloper, R.~Leigh, D.~Marolf, F.~Markopoulou, C.~Nappi,
J.~Polchinski,
L.~Smolin, A.~Schwarz, T.~Takeuchi, W.~Taylor
and R.~Zia for correspondence, comments and discussions.

\end{document}